\documentclass[showpacs,nofootinbib,titlepage,12pt,prd,preprint,groupedaddress,showkeys]{revtex4}
\usepackage{amsfonts}
\usepackage{amsmath}
\usepackage{amssymb}
\usepackage{graphicx}

\input{tcilatex}

\begin{document}

\title{Generalized dilaton-Maxwell cosmic string and wall solutions}
\author{John R. Morris}
\affiliation{Physics Dept., Indiana University Northwest, 3400 Broadway, Gary, Indiana
46408, USA}
\email{jmorris@iun.edu}

\begin{abstract}
The class of static solutions found by Gibbons and Wells for
dilaton-electrodynamics in flat spacetime, which describe nontopological
strings and walls that trap magnetic flux, is extended to a class of
dynamical solutions supporting arbitrarily large, nondissipative traveling
waves, using techniques previously applied to global and local topological
defects. These solutions can then be used in conjunction with S-duality to
obtain more general solitonic solutions for various axidilaton-Maxwell
theories. As an example, a set of dynamical solutions is found for axion,
dilaton, and Maxwell fields in low energy heterotic string theory using the $%
SL(2,\mathbb{R})$ invariance of the equations of motion.
\end{abstract}

\maketitle

\section{Introduction}

The common forms of cosmic string and domain wall solutions that have been
studied so extensively in the past\cite{VSbook} involve discrete or
continuous symmetries that are broken by the vacuum states. For instance, a
simple $Z_{2}$ domain wall described by a single real scalar field $\varphi $
arises from an action%
\begin{equation}
S_{W}=\int d^{4}x\sqrt{-g}\left[ \frac{1}{2}(\partial _{\mu }\varphi )^{2}-%
\frac{1}{4}\lambda (\varphi ^{2}-\eta ^{2})^{2}\right]  \label{e1}
\end{equation}

and a local $U(1)$ cosmic string described by the complex scalar field $\phi 
$ and gauge field $A_{\mu }$ can be obtained from an Abelian-Higgs action%
\begin{equation}
S_{CS}=\int d^{4}x\sqrt{-g}\left[ \frac{1}{2}(\left\vert D_{\mu }\phi
\right\vert ^{2}-\frac{1}{4}\lambda (\left\vert \phi \right\vert ^{2}-\eta
^{2})^{2}-\frac{1}{4}F_{\mu \nu }^{2}\right]  \label{e2}
\end{equation}

with $D_{\mu }=\partial _{\mu }+ieA_{\mu }$. The action for a global string
can be obtained by setting $A_{\mu }=0$. In the solutions admitted by these
actions the scalar self coupling $\lambda $ and the symmetry breaking
parameter $\eta $ determine the thickness of the wall or string. For the
local $U(1)$ string the magnetic flux is trapped inside the string core
where $\phi \rightarrow 0$ and the gauge field becomes massless. Whether or
not closed form analytic solutions are known for these solitonic objects, it
has been shown\cite{VV} that they can support traveling wave solutions of
arbitrary amplitude that are nondissipative, i.e., they do not result in the
radiation of scalar or gauge particles. Such radiation must come from
nontraveling wave solutions\cite{VEV}. The method used in ref.\cite{VV} to
show the existence of nondissipative traveling waves relies on the
definition of new coordinates $X$ and $Y$ that are related to the cartesian
coordinates $x$ and $y$ by $X=x-\psi (t,z)$, $Y=y-\chi (t,z)$, where $\psi
(t,z)$ and $\chi (t,z)$ are traveling wave solutions satisfying $(\partial
_{t}^{2}-\partial _{z}^{2})\xi (t,z)=0$, $\partial ^{\mu }\xi \partial _{\mu
}\xi =0$, where $\xi =\psi ,\chi $ and $\psi $ and $\chi $ are both
functions of either $t-z$ or $t+z$. Then the dynamical solutions, for
instance $\phi (t,x,y,z)$ and $A_{\mu }(t,x,y,z)$ for cosmic strings, are
obtained from the functions $\Phi (X,Y)$ and $\mathcal{A}_{\mu }(X,Y)$,
where $\Phi (x,y)$ and $\mathcal{A}_{\mu }(x,y)$ are static solutions for a
string stretching along the $z$ axis.

More specifically, for a static domain wall located in the $y-z$ plane, the
center of the wall is at at $x=0$. The replacement $x\rightarrow X=x-\psi
(t\pm z)$, $\Phi (x)\rightarrow \Phi (X)=\phi (x,z,t)$ gives a solution
describing a traveling wave in the wall, with the wall's center described by
the surface $X=x-\psi (t\pm z)=0$. In a similar manner, for a string
stretching along the $z$ axis, the replacements $x\rightarrow X=x-\psi (t\pm
z)$, $y\rightarrow Y=y-\chi (t\pm z)$ descibes a string with center at $X=0$%
, $Y=0$, and the dynamical solutions are given by\cite{VV,EHS} $\phi
(x,y,z,t)=\Phi (X,Y)$ and $A_{\mu }(x,y,z,t)=(\partial X^{M}/\partial x^{\mu
})\mathcal{A}_{M}(X,Y,z,t)$, where $X^{M}=(t,X,Y,z)$. Extended methods were
also developed to obtain the metric for gravitating solitons with traveling
waves\cite{GV,EHS}.

A 4d dilaton coupling to strings and walls (in the Einstein conformal frame)
can be included in an action of the form%
\begin{equation}
S^{\prime }=\int d^{4}x\sqrt{-g}\left\{ \frac{1}{2}(\partial _{\mu }\phi
)^{2}-U(\phi )+e^{-\alpha \phi }\mathcal{L}\right\}  \label{e3}
\end{equation}

where $\phi $ is a 4d dilaton field (which may be a modulus field associated
with a compactified extra dimension) and the Lagrangian $\mathcal{L}$ is
built from the scalar and gauge fields describing the wall or string\cite{GS}%
. (The Einstein-Hilbert contribution to the action has been dropped for the
case of a flat spacetime.) The dilaton field $\phi $ is a real valued scalar
field, but the fields giving rise to strings or walls are the fields in $%
\mathcal{L}$. Again, the strings or walls develop because of a broken
symmetry.

\section{Dilaton-Maxwell Theory}

The string and wall solutions found by Gibbons and Wells\cite{GW} are quite
different from the types of topological soliton solutions discussed above.
The Gibbons-Wells solutions are exact, analytical solutions of pure
dilaton-Maxwell theory. The action for dilaton-electrodynamics in a flat
spacetime is%
\begin{equation}
S=\int d^{4}x\left( \frac{1}{2}\partial _{\mu }\phi \partial ^{\mu }\phi -%
\frac{1}{4}e^{-2\tilde{\kappa}\phi }F_{\mu \nu }F^{\mu \nu }\right)
\label{e4}
\end{equation}

where $\tilde{\kappa}$ is a dimensioned parameter, not to be confused with a
Newtonian gravitation constant, and $g_{\mu \nu }=\eta _{\mu \nu }=(+,-,-,-)$%
. The action in (\ref{e4}) contains a real scalar field $\phi $ which
couples to the electromagnetic fields nonminimally. There is no spontaneous
symmetry breaking for $\phi $ as is the case for the ordinary string and
wall fields in the actions described in (\ref{e1}) -- (\ref{e3}).

For future convenience we list the electromagnetic field tensors%
\begin{equation}
F_{\mu \nu }=\left( 
\begin{array}{cccc}
0 & E_{x} & E_{y} & E_{z} \\ 
-E_{x} & 0 & -B_{z} & B_{y} \\ 
-E_{y} & B_{z} & 0 & -B_{x} \\ 
-E_{z} & -B_{y} & B_{x} & 0%
\end{array}%
\right) ,\ \ \ \ \ F^{\mu \nu }=\left( 
\begin{array}{cccc}
0 & -E_{x} & -E_{y} & -E_{z} \\ 
E_{x} & 0 & -B_{z} & B_{y} \\ 
E_{y} & B_{z} & 0 & -B_{x} \\ 
E_{z} & -B_{y} & B_{x} & 0%
\end{array}%
\right)  \label{e5a}
\end{equation}

with $E_{x}=E_{1}$, $B_{x}=B_{1}$, etc. being the physical fields and $%
F_{ij}=F^{ij}$. The equations of motion that follow from (\ref{e4}), along
with the Bianchi identity, are%
\begin{gather}
\square \phi +\frac{1}{2}\tilde{\kappa}e^{-2\tilde{\kappa}\phi }F_{\mu \nu
}F^{\mu \nu }=0  \label{e6} \\
\nabla _{\mu }\left( e^{-2\tilde{\kappa}\phi }F^{\mu \nu }\right) =0,\ \ \ \
\ \nabla _{\mu }\ \tilde{F}^{\mu \nu }=0  \label{e7}
\end{gather}

where the dual tensor is $\tilde{F}_{\mu \nu }=\frac{1}{2}\epsilon _{\mu \nu
\rho \sigma }F^{\rho \sigma }$. The set of equations in (\ref{e7}) is just
the set of Maxwell equations%
\begin{equation}
\nabla \cdot \mathbf{D}=0,\ \ \nabla \times \mathbf{H}-\mathbf{\dot{D}}=0,\
\ \nabla \cdot \mathbf{B}=0,\ \ \nabla \times \mathbf{E}+\mathbf{\dot{B}}=0
\label{e8}
\end{equation}

with $\mathbf{D}=\epsilon \mathbf{E}$ and $\mathbf{B}=\mu \mathbf{H}$ and
effective dielectric and permeability functions given by $\epsilon =\mu
^{-1} $, where%
\begin{equation}
\mu =\epsilon ^{-1}=e^{2\tilde{\kappa}\phi }  \label{e9}
\end{equation}

and the index of refraction is $\sqrt{\epsilon \mu }=1$. We can then rewrite
(\ref{e6}) as%
\begin{equation}
\nabla ^{2}\phi -\partial _{t}^{2}\phi =-\tilde{\kappa}e^{-2\tilde{\kappa}%
\phi }(\mathbf{B}^{2}-\mathbf{E}^{2})=-\tilde{\kappa}e^{2\tilde{\kappa}\phi
}(\mathbf{H}^{2}-\mathbf{D}^{2})  \label{e10}
\end{equation}

\section{Static Dilaton String and Wall Solutions}

The static $z$ - independent solutions discovered by Gibbons and Wells are
obtained by assuming an ansatz where $\phi =\Phi (x,y)$, $\mathbf{E}=0$, $%
\mathbf{H}=(0,0,\mathcal{H})=const$, and $\mathbf{B}=(0,0,\mathcal{B})=\mu 
\mathbf{H}$. I.e., we assume that there is a region of space where $%
H_{k}=\delta _{k3}\mathcal{H}$ is constant, so that%
\begin{equation}
\mathcal{\mathcal{B}}(x,y)=\mu (x,y)\mathcal{H}=e^{2\tilde{\kappa}\Phi (x,y)}%
\mathcal{H}  \label{e11}
\end{equation}

The Maxwell equations in (\ref{e8}) are then trivially satisfied, and the
dilaton equation (\ref{e10}) reduces to%
\begin{equation}
(\partial _{x}^{2}+\partial _{y}^{2})\Phi =-\tilde{\kappa}\mathcal{H}^{2}e^{2%
\tilde{\kappa}\Phi }  \label{e12}
\end{equation}

Equation (\ref{e12}) is the euclidean Liouville equation\cite{Liou} (see
also, for example, refs\cite{DJ,Witten}). Its solution\cite{GW} can be
written in terms of $\zeta =x+iy$ as%
\begin{equation}
\mu (\zeta )=e^{2\tilde{\kappa}\Phi (\zeta )}=\frac{4}{\tilde{\kappa}^{2}%
\mathcal{H}^{2}}\frac{\left\vert f^{\prime }(\zeta )\right\vert ^{2}}{\left(
1+\left\vert f(\zeta )\right\vert ^{2}\right) ^{2}}  \label{e13}
\end{equation}

where $f$ is a holomorphic function of $\zeta $ and $f^{\prime }(\zeta
)=df(\zeta )/d\zeta $. As pointed out in ref.\cite{GW}, the functions $f$
and $1/f$ give the same solution $\Phi $.

The interesting feature of the class of solutions given by (\ref{e11}) and (%
\ref{e13}) is that there is a trapping of magnetic flux in the solution
cores where $\Phi $, $\mu $, and $\mathcal{B}$ all become large. Gibbons and
Wells present two specific examples of solutions, one for a cylindrically
symmetric cosmic string, and one for a planar wall.

A cylindrically symmetric solution describing a cosmic string lying along
the $z$ axis is given by $f=a\zeta $ with $a$ being a constant. The solution
for $\Phi $ is then%
\begin{equation}
\mu =e^{2\tilde{\kappa}\Phi }=\frac{4a^{2}}{\tilde{\kappa}^{2}\mathcal{H}^{2}%
}\frac{1}{\left( 1+a^{2}\rho ^{2}\right) ^{2}}  \label{e14}
\end{equation}

where $\rho ^{2}=\left\vert \zeta \right\vert ^{2}=x^{2}+y^{2}$. The total
magnetic flux%
\begin{equation}
\Phi _{mag}=\int \mathbf{B}\cdot \hat{n}dA=\pi \mathcal{H}\int_{0}^{\infty
}\mu (\rho )d(\rho ^{2})=\frac{4\pi }{\tilde{\kappa}^{2}\mathcal{H}}
\label{e15}
\end{equation}

associated with this solution is finite with the flux being concentrated
near the core. The parameter $a$ is arbitrary and the total flux does not
depend on it. (Around 80\% of the total flux is trapped within a region
where $\rho a\lesssim 2.5$.) We also have that $\mu $, $\mathcal{B}$ $%
\rightarrow 0$ as $\rho \rightarrow \infty $. The string solution has a
logarithmically divergent energy per unit length, resembling that of a
global string rather than a local one.

For a wall-like solution Gibbons and Wells consider a solution generated by $%
f(\zeta )=\exp (\zeta )$, giving%
\begin{equation}
\mu =e^{2\tilde{\kappa}\Phi }=\frac{1}{\tilde{\kappa}^{2}\mathcal{H}^{2}}%
\frac{1}{\cosh ^{2}x}  \label{e16}
\end{equation}

This describes a wall-like solution with an amount of flux per unit length
of $\frac{2}{\tilde{\kappa}^{2}\mathcal{H}}$. Inside the wall $\mu $, $%
\mathcal{B}$ are large, and asymptotically $\mu $, $\mathcal{B}$ $%
\rightarrow 0$ as $x\rightarrow \pm \infty $.

\section{Solutions with Traveling Waves}

We use the solution generating technique developed by Vachaspati and
Vachaspati \cite{VV} to describe traveling waves on strings and walls. We
want to find solutions $\phi (x,y,z,t),F_{\mu \nu }(x,y,z,t)$ to the
equations of motion (\ref{e8}) -- (\ref{e10}). We start with static
solutions described by (\ref{e11}) -- (\ref{e13}) and make replacements $%
x\rightarrow X$, $y\rightarrow Y$; that is, we make a coordinate
transformation from the coordinates $x^{\mu }=(t,x,y,z)$ to the coordinates $%
X^{M}$, where we define the new coordinates%
\begin{eqnarray}
X^{M} &=&(X^{0},X^{1},X^{2},X^{3})=(t,X,Y,z),  \notag \\
X &=&x-\psi (t,z),\ \ \ Y=y-\chi (t,z),  \label{e17} \\
X^{0} &=&x^{0}=t,\ \ \ X^{3}=x^{3}=z  \notag
\end{eqnarray}

The functions $\psi $ and $\chi $ must be traveling waves that satisfy $%
(\partial _{t}^{2}-\partial _{z}^{2})\xi =0$ and $\partial _{\mu }\xi
\partial ^{\mu }\xi =0$, where $\xi =\psi ,\chi $ so that both $\psi $ and $%
\chi $ are functions of either $t-z$ or $t+z$. We then take the dilaton
field to be given by%
\begin{equation}
\phi (x^{\mu })=\phi (x,y,z,t)=\Phi (X,Y)  \label{e17a}
\end{equation}

Since $(\partial _{t}^{2}-\partial _{z}^{2})\Phi (X,Y)=0$, we have $\Phi
(X,Y)$ satisfying an equation of motion having the same form as in (\ref{e12}%
) (with the same constant $\mathcal{H}$) with the replacements $x\rightarrow
X$ and $y\rightarrow Y$. Of course, we must also determine the new $\mathbf{E%
}$ and $\mathbf{B}$ fields for the new coordinate system so that we get the
same right hand side of (\ref{e10}) with $e^{2\tilde{\kappa}\phi }(\mathbf{H}%
^{2}-\mathbf{D}^{2})=e^{2\tilde{\kappa}\Phi (X,Y)}\mathcal{H}^{2}$. This is
easily done by considering the tensor transformation of the field tensor $%
\mathcal{F}_{MN}(X^{M})$, where $\mathcal{F}_{MN}$ is composed of the
electric and magnetic fields%
\begin{equation}
\mathcal{\mathcal{E}}_{k}=0,\ \ \ \ \mathcal{\mathcal{B}}_{k}(X,Y)=\delta
_{k3}\mathcal{B}(X,Y)=\delta _{k3}e^{2\tilde{\kappa}\Phi (X,Y)}\mathcal{H}
\label{e18}
\end{equation}%
That is, we use the static solutions with $x,y\rightarrow X,Y$. We then have%
\begin{equation}
F_{\mu \nu }(x^{\alpha })=\frac{\partial X^{M}}{\partial x^{\mu }}\frac{%
\partial X^{N}}{\partial x^{\nu }}\mathcal{F}_{MN}(X^{A})  \label{e19}
\end{equation}

Using (\ref{e5a}), (\ref{e18}) and (\ref{e19}) give the dynamical electric
and magnetic fields $\vec{E}(x,y,z,t)$ and $\vec{B}(x,y,z,t)$ with components%
\begin{equation}
\begin{array}{ll}
E_{1}=E_{x}=-\dot{\chi}\mathcal{B} & B_{1}=B_{x}=\psi ^{\prime }\mathcal{B}
\\ 
E_{2}=E_{y}=\dot{\psi}\mathcal{B} & B_{2}=B_{y}=\chi ^{\prime }\mathcal{B}
\\ 
E_{3}=E_{z}=(\psi ^{\prime }\dot{\chi}-\dot{\psi}\chi ^{\prime })\mathcal{B}%
=0\ \ \ \  & B_{3}=B_{z}=\mathcal{B}%
\end{array}
\label{e20}
\end{equation}

where $\dot{\psi}=\partial _{0}\psi ,$ $\psi ^{\prime }=\partial _{z}\psi ,$
etc. We also have for the fields $\mathbf{D(}x^{\mu })$ and $\mathbf{H(}%
x^{\mu })$ 
\begin{subequations}
\label{e21}
\begin{eqnarray}
\mathbf{D} &=&\epsilon \mathbf{E}=e^{-2\tilde{\kappa}\Phi (X,Y)}\mathbf{E}
\label{e21a} \\
\mathbf{B} &=&\mu \mathbf{H}=e^{2\tilde{\kappa}\Phi (X,Y)}\mathbf{H}
\label{e21b}
\end{eqnarray}

From $\mathbf{H}=\mu ^{-1}\mathbf{B}$ and $B_{3}=\mathcal{B}=e^{2\tilde{%
\kappa}\Phi (X,Y)}\mathcal{H}$, we find $H_{3}=H_{z}=\mathcal{H}$ and $e^{2%
\tilde{\kappa}\phi }(\mathbf{H}^{2}-\mathbf{D}^{2})=e^{2\tilde{\kappa}\Phi
(X,Y)}\mathcal{H}^{2}$.

Using the facts that $\partial _{x}=\partial _{X}$, $\partial _{y}=\partial
_{Y}$, one can verify that the solutions given by (\ref{e17a}) and (\ref{e20}%
) are $x,y,z,t$ dependent solutions to the equations of motion (\ref{e8}) --
(\ref{e10}). Note that we \textit{do not} make a tensor transformation on
the metric, but rather we take $g_{\mu \nu }(x^{\mu })=g_{MN}(X^{M})$, i.e.,
we take $\eta _{\mu \nu }=\eta _{MN}$ in order to maintain the \textit{form
invariance} of the equations of motion, as opposed to a covariance under a
general coordinate transformation. The solutions $\phi (x^{\mu }),$ $F_{\mu
\nu }(x^{\mu })$ describe traveling waves of arbitrary size propagating in
the $\pm z$ direction through otherwise static string or wall solutions that
trap magnetic flux. The center of a wall is now located at $X=x-\psi (t,z)=0$%
, and the center of a string at $X=x-\psi (t,z)=0$, $Y=y-\chi (t,z)=0$, and
the entrapped region of magnetic flux in the core of the solution moves
along with the dilaton field. Furthermore, since the nonstatic field field $%
\Phi (X,Y)$ has the same functional form as the static field $\Phi (x,y)$
and the elctromagnetic fields $\mathbf{E}$ and $\mathbf{B}$ are proportional
to $\mathcal{B}(X,Y)$,and $\mathcal{B}(X,Y)$ has the same functional form as
the static field $\mathcal{B}(x,y)$, then if the static solutions drop off
to zero away from the solution core, then the same must hold for the
nonstatic solutions. We therefore conclude that the dynamical solutions are
nondissipative and do not radiate energy in the form of either dilatons or
photons.

To make this more explicit, we note that the scalar dilaton field can be
written in terms of a traveling wave part $\Phi (X,Y)$ plus a radiative part 
$\delta \phi (x,y,z,t)$, i.e., $\phi (x,y,z,t)=\Phi (X,Y)+\delta \phi
(x,y,z,t)$, as done for the case of scalar radiation from oscillating domain
walls in ref.\cite{VEV}. Since $\delta \phi =0$ for the case of traveling
waves, these solutions generate no dilatonic radiation in the form of
particles escaping to an infinite distance from the soliton core. We
conclude that scalar dilaton radiation can come only from solitons that
support nontraveling waves (e.g., standing waves). Concerning
electromagnetic radiation, we note that the motion of the soliton core will
be associated with a locally time dependent magnetic flux and induced
electric field, so that there is, in general, a locally nonvanishing
Poynting vector. From the fields in (\ref{e20}) and (\ref{e21}) we find 
\end{subequations}
\begin{equation}
\vec{S}=\vec{E}\times \vec{H}=\epsilon (X,Y)(\vec{E}\times \vec{B})=\mathcal{%
HB}(X,Y)\left( \dot{\psi},\dot{\chi},-(\dot{\psi}\psi ^{\prime }+\dot{\chi}%
\chi ^{\prime })\right)  \label{e22}
\end{equation}

The equation of motion for the dilaton can be written in the form%
\begin{equation}
\nabla ^{2}\Phi =-\tilde{\kappa}\mathcal{HB}  \label{e23}
\end{equation}

For a single localized soliton with $\nabla \Phi \rightarrow 0$
asymptotically, we must have that $\mathcal{B}\rightarrow 0$ at asymptotic
distances, which implies that $\vec{S}\rightarrow 0$ at asymptotic distances
from the solitonic core. To have $\int \vec{S}\cdot \hat{n}dA\rightarrow 0$
through a surface asymptotically distant from the soliton, we require that $%
\mathcal{B}\rightarrow 0$ faster than the inverse distance, asymptotically.
For example, if asymptotically $\left\vert \nabla \Phi \right\vert \sim \rho
^{-n}$, $n>0$, for a string-like solution, then $\left\vert \nabla ^{2}\Phi
\right\vert \sim \rho ^{-(n+1)}\sim \mathcal{B}$. It follows that $S_{\rho }=%
\mathcal{HB}(X,Y)\left( \frac{x}{\rho }\dot{\psi}+\frac{y}{\rho }\dot{\chi}%
\right) \sim \mathcal{B}\rightarrow 0$ faster than $\rho ^{-1}$, so that no
electromagnetic energy flows radially in or out through a cylindrical
surface at $\rho \rightarrow \infty $. A similar argument can be made for a
wall-like solution. However, electromagnetic energy does flow in and out
from the soliton core \textit{locally} as the soliton wiggles around. Also
there is an energy flow in the $z$ direction, within the string or wall.
Also, multisoliton solutions (e.g., a network of strings\cite{GW}) exist,
and each soliton is assumed to have the same type of radiative behavior as a
single isolated soliton.

Finally, we can note that the gauge field $A_{\mu }(x^{\mu })$ for the
dynamical solutions can be related to the gauge field $\mathcal{A}%
_{M}(X^{M}) $ for the static solutions, with $x^{\mu }\rightarrow X^{M}$, by
the tensor transformation\cite{EHS}%
\begin{equation}
A_{\mu }(x^{\sigma })=(\partial _{\mu }X^{M})\mathcal{A}_{M}(X^{S})
\label{e24}
\end{equation}

where $X^{M}$ is given by (\ref{e17}) and $A_{\mu }(x)$ and $\mathcal{A}%
_{M}(X)$ generate the field strengths $F_{\mu \nu }(x)$ and $\mathcal{F}%
_{MN}(X)$, respectively. From $F_{\mu \nu }(x)=\partial _{\mu }A_{\nu
}(x)-\partial _{\nu }A_{\mu }(x)$ and $\mathcal{F}_{MN}(X)=\partial _{M}%
\mathcal{A}_{N}(X)-\partial _{N}\mathcal{A}_{M}(X)$, (\ref{e24}) yields (\ref%
{e19}), the field tensor for the dynamical solutions.

\section{Extensions Using S-Duality}

The equations of motion given by (\ref{e6}) and (\ref{e7}) exhibit an
S-duality, i.e., they are invariant under the duality transformations%
\begin{equation}
\phi \rightarrow -\phi ,\ \ \ \ F_{\mu \nu }\rightarrow e^{-2\tilde{\kappa}%
\phi }\ \tilde{F}_{\mu \nu }  \label{e25}
\end{equation}

where $\tilde{F}_{\mu \nu }=\frac{1}{2}\epsilon _{\mu \nu \rho \sigma
}F^{\rho \sigma }$. This symmetery could be used to generate new solutions.
For example, if the duality transformations are applied to the static
solutions of (\ref{e11}) and (\ref{e13}), we have $\epsilon \rightarrow \mu
, $ $\mu \rightarrow \epsilon ,$ $\mathbf{E}\rightarrow \mathbf{H},$ $%
\mathbf{B}\rightarrow -\mathbf{D},$ $\mathbf{D}\rightarrow \mathbf{B},$ and $%
\mathbf{H}\rightarrow -\mathbf{E}$. Consequently we have the new solutions 
\begin{subequations}
\label{e26}
\begin{eqnarray}
\mathcal{\mathcal{D}}(x,y) &=&\epsilon (x,y)\mathcal{\mathcal{E}}=e^{-2%
\tilde{\kappa}\Phi (x,y)}\mathcal{\mathcal{E}}  \label{e26a} \\
\epsilon (\zeta ) &=&e^{-2\tilde{\kappa}\Phi (\zeta )}=\frac{4}{\tilde{\kappa%
}^{2}\mathcal{\mathcal{E}}^{2}}\frac{\left\vert f^{\prime }(\zeta
)\right\vert ^{2}}{\left( 1+\left\vert f(\zeta )\right\vert ^{2}\right) ^{2}}
\label{e26b}
\end{eqnarray}

describing strings or walls in a region of constant electric field $\mathbf{E%
}=(0,0,\mathcal{E})$ that entrap the displacement field $\mathcal{D}$, as
pointed out in\cite{GW}.

This method can also be used in effective field theories with more fields
and other dual symmetries, such as $SL(2,\mathbb{R})$. An example is
provided in\cite{DJM} where the authors consider an effective 4d low energy
heterotic string theory containing electromagnetic, dilaton, and axion
fields. Starting with a solution with zero axion, nontrivial new solutions
with a nonzero axion can be generated from the $SL(2,\mathbb{R})$ symmetry.
This technique could be applied to other theories exhibiting $SL(2,\mathbb{R}%
)$ duality as well, such as supergravities\cite{LO} and nonlinear
electrodynamics coupled to an axion and dilaton\cite{GR1,GR2}, and to low
energy string theories with generalized couplings of dilaton and modulus
fields to gauge fields, along with generalized duality transformations\cite%
{CT}. Here, we follow the example in ref.\cite{DJM} by taking the dynamic
Gibbons-Wells type of solutions obtained in the last section, along with a
vanishing axion field, and using these solutions as input to generate, via
the $SL(2,\mathbb{R})$ dual symmetry, new dynamical solutions for an
effective low energy heterotic string theory that possesses a nonzero axion
field (in the flat space limit). Using the units of ref.\cite{DJM}, we write
the action (in a flat spacetime) for a truncated version of the effective 4d
low energy heterotic string theory\cite{DJM} as\footnote{%
We use a different (mostly negative) metric.} 
\end{subequations}
\begin{equation}
S=\int d^{4}x\left[ \frac{1}{2}(\partial \phi )^{2}+\frac{1}{12}e^{-2\phi
}H_{\mu \nu \rho }H^{\mu \nu \rho }-e^{-\phi }F_{\mu \nu }F^{\mu \nu }\right]
\label{e27}
\end{equation}

where $\phi $ is the dilaton, $H_{\mu \nu \rho }$ is an antisymmetric rank
three tensor field, and the axion field $\Psi $ is defined through%
\begin{equation}
H^{\mu \nu \rho }=-e^{2\phi }\epsilon ^{\mu \nu \rho \sigma }\partial
_{\sigma }\Psi  \label{e28}
\end{equation}

The equations of motion are 
\begin{subequations}
\label{e30}
\begin{gather}
\nabla _{\mu }(e^{-\phi }F^{\mu \nu }+\Psi \tilde{F}^{\mu \nu })=0
\label{e30a} \\
\nabla _{\mu }\tilde{F}^{\mu \nu }=0  \label{e30b} \\
\square \phi -e^{-\phi }F_{\mu \nu }F^{\mu \nu }-e^{2\phi }\partial _{\mu
}\Psi \partial ^{\mu }\Psi =0  \label{e30c} \\
\square \Psi +2\partial _{\mu }\phi \partial ^{\mu }\Psi +e^{-2\phi }F_{\mu
\nu }\tilde{F}^{\mu \nu }=0  \label{e30d}
\end{gather}

Upon setting $\Psi =0,$ $\mathbf{E}=0,$ $B_{k}=\delta _{3k}\mathcal{B}$,
these equations reduce to the form of those in (\ref{e6}) and (\ref{e7}),
the solutions to which are given by (\ref{e17a}) and (\ref{e19}). The
equations of motion in (\ref{e30}) are invariant under the duality
transformations\cite{DJM} 
\end{subequations}
\begin{equation}
\lambda \rightarrow \lambda ^{\prime }=\frac{a\lambda +b}{c\lambda +d},\ \ \
\ F_{\mu \nu }\rightarrow F_{\mu \nu }^{\prime }=(c\Psi +d)F_{\mu \nu
}-ce^{-\phi }\tilde{F}_{\mu \nu },\ \ \ \ ad-bc=1  \label{e31}
\end{equation}

where $\lambda =\Psi +ie^{-\phi }$. Using the solutions in (\ref{e17a}) and (%
\ref{e19}) along with $\Psi =0$ as input, (\ref{e31}) yields a new solution
set given by%
\begin{equation}
\Psi ^{\prime }=\frac{ace^{-2\phi }+bd}{c^{2}e^{-2\phi }+d^{2}},\ \ \ \
e^{-\phi \prime }=\frac{e^{-\phi }}{c^{2}e^{-2\phi }+d^{2}},\ \ \ \ F_{\mu
\nu }^{\prime }=dF_{\mu \nu }-ce^{-\phi }\tilde{F}_{\mu \nu }  \label{e32}
\end{equation}

with the various components of $F_{\mu \nu }$ being given by (\ref{e20}).
This represents a generalization of the class of solutions of the
Gibbons-Wells type describing wiggly strings and walls, with nonvanishing
axion fields, for the effective low energy heterotic string theory described
by (\ref{e27}). These types of solutions may be of physical interest, for
example, in providing a natural solution to the monopole problem\cite{GW}.

\section{Brief Summary}

To summarize, we have found dynamical solutions of the flat-space
dilaton-Maxwell theory that describe nontopological solitons entrapping
magnetic flux and supporting nondissipative traveling waves. These
solutions, given by eqs. (\ref{e17a})--(\ref{e21}), depend on a holomorphic
function $f(\boldsymbol{Z})$, where $\boldsymbol{Z}=X+iY$, and%
\begin{equation}
\mu (\boldsymbol{Z})=e^{2\tilde{\kappa}\Phi (\boldsymbol{Z})}=\frac{4}{%
\tilde{\kappa}^{2}\mathcal{H}^{2}}\frac{\left\vert f^{\prime }(\boldsymbol{Z}%
)\right\vert ^{2}}{\left( 1+\left\vert f(\boldsymbol{Z})\right\vert
^{2}\right) ^{2}}  \label{e33}
\end{equation}

These $z$ independent solitons can represent codimension 1 or 2 objects, or
a network of such objects, depending upon the choice for the function $f$.
These static or dynamic solutions, along with their dilaton-Maxwell duals
(see, e.g., (\ref{e26})), can be further used as seed solutions in
axidilaton-Maxwell theories with the use of S-duality. An example has been
provided for low energy string theory with toroidal compactification, where
the dilaton-Maxwell solutions are used along with a vanishing axion field to
generate a new class of $SL(2,\mathbb{R})$ dual solutions describing
different static or dynamic solitonic flux-trapping configurations with
nonvanishing axion field (eq.(\ref{e32})). The types of nontopological
solitons discussed here differ from the usual topological solitons not only
from a mathematical standpoint, but also have different physical
characteristics, and can arise naturally in classes of axidilaton-Maxwell
theories, including low energy string theories.

\end{document}